\pdfoutput=1

\documentclass[11pt,twoside,a4paper,cmspaper,final,collab]{cms-tdr}
\def\svnVersion{439406P}\def\cmsCernNoTag{CERN-EP-2017-196}\def\cmsCernDate{\today}\def\cmsMessage{Published in Physics Letters B as \href{http://dx.doi.org/10.1016/j.physletb.2017.12.006}{\doi{10.1016/j.physletb.2017.12.006.}}}

\begin{document}\cmsNoteHeader{B2G-17-010}

\hyphenation{had-ron-i-za-tion}
\hyphenation{cal-or-i-me-ter}
\hyphenation{de-vices}
\RCS$Revision: 439399 $
\RCS$HeadURL: svn+ssh://svn.cern.ch/reps/tdr2/papers/B2G-17-010/trunk/B2G-17-010.tex $
\RCS$Id: B2G-17-010.tex 439399 2017-12-18 19:13:49Z drankin $
\newlength\cmsFigWidth
\ifthenelse{\boolean{cms@external}}{\setlength\cmsFigWidth{0.85\columnwidth}}{\setlength\cmsFigWidth{0.4\textwidth}}
\ifthenelse{\boolean{cms@external}}{\providecommand{\cmsLeft}{top\xspace}}{\providecommand{\cmsLeft}{left\xspace}}
\ifthenelse{\boolean{cms@external}}{\providecommand{\cmsRight}{bottom\xspace}}{\providecommand{\cmsRight}{right\xspace}}
\providecommand{\cmsTable}[1]{\resizebox{\textwidth}{!}{#1}}
\newcommand{\mtb}{\ensuremath{M_{\PQt\PQb}}\xspace}
\newcommand{\mc}[3]{\multicolumn{#1}{#2}{#3}}

\cmsNoteHeader{B2G-17-010}
\title{Search for heavy resonances decaying to a top quark and a bottom quark in the lepton+jets final state in proton-proton collisions at 13\TeV}

\date{\today}

\abstract{
A search is presented for narrow heavy resonances decaying to a top quark and a bottom quark using data collected by the CMS experiment at $\sqrt{s}=13\TeV$ in 2016. The data set analyzed corresponds to an integrated luminosity of 35.9\fbinv. Final states that include a single lepton ($\Pe,\PGm$), multiple jets, and missing transverse momentum are analyzed. No evidence is found for the production of a $\PWpr$ boson, and the production of right-handed $\PWpr$ bosons is excluded at 95\% confidence level for masses up to 3.6\TeV depending on the scenario considered. Exclusion limits for $\PWpr$ bosons are also presented as a function of their coupling strength to left- and right-handed fermions. These limits on a $\PWpr$ boson decaying via a top and a bottom quark are the most stringent published to date.
}

\hypersetup{%
pdfauthor={CMS Collaboration},%
pdftitle={Search for heavy resonances decaying to a top quark and a bottom quark in the lepton+jets final state in proton-proton collisions at 13 TeV},%
pdfsubject={CMS},%
pdfkeywords={CMS, physics, B2G, exotica, Wprime}
}

\maketitle

\section{Introduction}

Despite the broad success of the standard model (SM), the absence of answers to the hierarchy problem, among other shortcomings, has led to the development of many theories for new physics that lies beyond the SM.
A common prediction of many of these theories is the existence of new heavy gauge bosons~\cite{doi:10.1146/annurev.nucl.55.090704.151502,PhysRevD.64.035002,PhysRevD.64.065007,PhysRevD.53.5258,PhysRevD.11.566}.
These particles typically arise from additional symmetries in the theories, and it is common to generically refer to charged instances of these resonances as $\PWpr$ bosons.
In scenarios where the $\PWpr$ boson is sufficiently heavy, the decay $\PWpr\to\PQt\PQb$ has several features that make it an appealing search channel.
Searches in this channel directly probe the $\PWpr$ boson coupling to third generation quarks, which, in some models~\cite{Muller1996345,Malkawi1996304}, can be enhanced with respect to the coupling to lighter quarks.
Additionally, the large continuum multijet background has less impact on searches for $\PWpr\to\PQt\PQb$ decay than on searches for the decay to light quarks ($\PWpr\to \PQq\PQq'$).
The  $\PWpr\to\PQt\PQb$ search is complementary to searches for $\PWpr\to{\ell}\PGn$ and $\PWpr\to\PW\Z$, where $\ell$ denotes a charged lepton and $\PGn$ denotes a neutrino.
Unlike searches for $\PWpr\to\ell\PGn$, the search for $\PWpr\to\PQt\PQb\to\PQb\PQb\ell\PGn$ decay allows the $\PWpr$ boson mass to be fully reconstructed, up to a quadratic ambiguity.

Searches for  $\PWpr$ bosons in the top and bottom quark (tb) decay channel have been performed at the Fermilab Tevatron~\cite{D0:2010,D0Wprime,PhysRevLett.115.061801} and at the CERN LHC by both CMS~\cite{Chatrchyan:2014koa,Chatrchyan:2016had,Sirunyan:2017ukk} and ATLAS~\cite{Aad:2014xra,Aad:2014xea} Collaborations.
The most stringent limits to date on the production of $\PWpr$ bosons come from the CMS search performed at $\sqrt{s}=13\TeV$~\cite{Sirunyan:2017ukk}, using 2.2\fbinv of data collected in 2015.

This Letter presents a search for $\PWpr$ bosons decaying via the tb channel using proton-proton collision data at $\sqrt{s}=13\TeV$, collected by the CMS experiment in 2016.
The analyzed data correspond to an integrated luminosity of 35.9\fbinv.
Events with exactly one electron or muon, significant missing transverse momentum, and multiple jets in the final state are selected.
This search focuses on $\PWpr$ bosons with widths that are narrow compared to their masses.
In addition to searching for $\PWpr$ bosons with purely right- or left-handed couplings, we also search for $\PWpr$ bosons with varying combinations of these couplings.
This analysis is sensitive to $\PWpr$ bosons with masses between 1 and 4\TeV.

\section{The CMS detector\label{sec:detector}}

The central feature of the CMS apparatus~\cite{Chatrchyan:2008zzk} is a superconducting solenoid of 6\unit{m} internal diameter, providing a magnetic field of 3.8\unit{T}.
Within the solenoid volume are a silicon pixel and strip tracker, a lead tungstate crystal electromagnetic calorimeter (ECAL), and a brass and scintillator hadron calorimeter (HCAL), each composed of a barrel and two endcap sections.
Forward calorimeters extend the pseudorapidity ($\eta$) coverage provided by the barrel and endcap detectors.
Muons are measured in gas-ionization detectors embedded in the steel flux-return yoke outside the solenoid.

The particle-flow (PF) algorithm~\cite{Sirunyan:2017ulk} reconstructs and identifies individual particle candidates with an optimized combination of information from relevant elements of the CMS detector.
The energy of photons is measured using the ECAL and corrected for zero-suppression effects.
The energy of electrons is determined from a combination of the electron momentum at the primary interaction vertex as determined by the tracker, the energy of the corresponding ECAL cluster, and the energy sum of all bremsstrahlung photons spatially compatible with originating from the electron track.
The primary interaction vertex is defined as the vertex with the largest sum of $\pt^2$ of associated tracks.
The energies of muons are obtained from the curvature of the corresponding tracks.
The energy of charged hadrons is determined from a combination of their momentum measured in the tracker and the matching ECAL and HCAL energy deposits.
This measurement is then corrected for zero-suppression effects and for the response function of the calorimeters to hadronic showers.
Finally, the energy of neutral hadrons is obtained from the corresponding corrected ECAL and HCAL energy.

In the barrel section of the ECAL, an energy resolution of about 1\% is achieved for unconverted or late-converting photons in the tens of \GeV energy range.
The resolution for photons not belonging to this category is about 1.3\% up to $\abs{\eta} = 1$, rising to about 2.5\% at $\abs{\eta} = 1.4$.
In the endcaps, the resolution of unconverted or late-converting photons is about 2.5\%, while the remaining photons have a resolution between 3 and 4\%~\cite{CMS:EGM-14-001}.
The momentum resolution for electrons with transverse momentum $\pt\approx45$\GeV from $\Z \to \Pe \Pe$ decays ranges from 1.7\% for nonshowering electrons in the barrel region to 4.5\% for showering electrons in the endcaps~\cite{Khachatryan:2015hwa}.
When combining information from the entire detector, the jet energy resolution amounts typically to 15\% at 10\GeV, 8\% at 100\GeV, and 4\% at 1\TeV, to be compared to about 40, 12, and 5\% obtained when the ECAL and HCAL calorimeters alone are used~\cite{Chatrchyan:2011ds}.

Muons are measured in the range $\abs{\eta} < 2.4$, with detection planes made using three technologies: drift tubes, cathode strip chambers, and resistive-plate chambers.
Matching muons to tracks measured in the silicon tracker results in a relative transverse momentum resolution for muons with $20 <\pt < 100\GeV$ of 1.3--2.0\% in the barrel, and better than 6\% in the endcaps.
The \pt resolution in the barrel is better than 10\% for muons with \pt up to 1\TeV~\cite{Chatrchyan:2012xi}.

The missing transverse momentum vector \ptvecmiss is defined as the projection on the plane perpendicular to the beams of the negative vector sum of the momenta of all reconstructed particles in an event.

A more detailed description of the CMS detector, together with a definition of the coordinate system used and the relevant kinematic variables, can be found in Ref.~\cite{Chatrchyan:2008zzk}.
\section{Signal and background modeling\label{sec:model}}

\subsection{Signal modeling}

Simulated signal samples are generated at leading order and their cross sections are scaled to next-to-leading order with a K-factor of 1.25 \cite{Sullivan:2002jt,Duffty:2012rf} appropriate for our signal mass range of interest.
All signal samples are generated using the \COMPHEP \cite{comphep} 4.5.2 package according to the following lowest-order effective Lagrangian~\cite{Sullivan:2002jt}:
\begin{equation}
 \mathcal{L}=\frac{V_{f_if_j}}{2\sqrt{2}}g_{\PW}\bar{f}_i\gamma_\mu \left[a_\mathrm{R}(1+\gamma^5)+a_\mathrm{L}(1-\gamma^5)\right]\mathrm{W}'^\mu f_j+\text{h.c.},\label{eq:lag}
\end{equation}
where $V_{f_if_j}$ is the Cabibbo--Kobayashi--Maskawa matrix if $f$ is a quark and $V_{f_if_j}=\delta_{ij}$ if $f$ is a lepton, $g_{\PW}$ is the SM weak coupling constant, and $a_\mathrm{R}$ and $a_\mathrm{L}$ are the coupling strengths of the \PWpr to right- and left-handed fermions, respectively.
We consider values of $a_\mathrm{L}$ and $a_\mathrm{R}$ that range from 0 to 1, and any signal with $a_\mathrm{L}>0$ takes into account interference with the SM W boson.
The signal simulation includes decays involving a $\tau$ lepton,
and no distinction is made in the analysis selection or strategy between an electron or muon produced directly from the W boson decay, and an electron or muon from a subsequent $\tau$ lepton decay.
We use \PWpr boson width values computed in \COMPHEP for each mass point, and use a narrow-width approximation for the generation of \PWpr bosons that have both left- and right-handed couplings.
The typical width is approximately 3\% of the signal resonance mass.
The widths of all generated samples are significantly smaller than the detector and reconstruction resolutions, and therefore the precise values of the width do not affect our results.

For $\PWpr_\mathrm{R}$ bosons we consider two scenarios for the mass of the hypothetical right-handed neutrinos.
If the right-handed neutrinos are lighter than the $\PWpr_\mathrm{R}$ boson ($M_{\nu_\mathrm{R}}<M_{\PWpr_\mathrm{R}}$), then both $\PWpr_\mathrm{R}\to\ell\nu_\mathrm{R}$ and $\PWpr_\mathrm{R}\to\PQq\PQq'$ decays are allowed.
However, if the right-handed neutrinos are heavier than the $\PWpr_\mathrm{R}$ boson ($M_{\nu_\mathrm{R}}>M_{\PWpr_\mathrm{R}}$), then the $\PWpr_\mathrm{R}\to\ell\nu_\mathrm{R}$ decay is forbidden, resulting in an enhancement of the branching fraction for $\PWpr\to\PQt\PQb$.
This branching fraction varies slightly with mass and ranges from 0.32 to 0.33 if $M_{\nu_\mathrm{R}}>M_{\PWpr_\mathrm{R}}$ and from 0.24 to 0.25 if $M_{\nu_\mathrm{R}}<M_{\PWpr_\mathrm{R}}$ for $\PWpr_\mathrm{R}$ boson masses between 1 and 4\TeV.
For the purposes of signal generation all neutrinos are assumed to be massless.
When calculating the number of expected signal events (in Table~\ref{table:lep_yields}), showing expected signal distributions (in Figs.~\ref{fig:lep_mtb_A} and~\ref{fig:lep_mtb_B}), or presenting results for arbitrary left- and right-handed couplings (in Fig.~\ref{fig:2dlimit}), it is always assumed
that the masses of hypothetical right-handed neutrinos are much lighter than that of the $\PWpr_\mathrm{R}$ boson.
Both scenarios are considered when presenting results for $\PWpr_\mathrm{R}$ (in Figs.~\ref{fig:sepxseclim_sep} and~\ref{fig:sepxseclim}).

\subsection{Background modeling}

The most significant contributions to the background come from W+jets and \ttbar production.
Smaller contributions, from $s$- and $t$-channel single top quark production, associated production of a top quark and a W boson, $\Z/\gamma^*$+jets, and diboson production ($\mathrm{VV}$), are also included in the total background estimate.
Predictions for all background processes are taken from simulation with corrections applied in cases where initial modeling is found to be inaccurate.
Further details on the background modeling can be found in Section~\ref{sec:anstrat}.
The contribution to the total background from the multijet background is found to be negligible after the full selection and is therefore not included.

Simulated samples for $\Z/\gamma^*$+jets, $s$- and $t$-channel single-top quark, and W+jets events are produced using \MGvATNLO~\cite{Alwall:2014hca,Frederix:2012ps,Alwall:2007fs} v2.2.2, \ttbar and associated production of a top quark and a W boson are produced using \POWHEG v2 ~\cite{Nason:2004rx,powheg,Alioli:2010xd,Frixione:2007nw,Re:2010bp}, and all other background processes are produced using \PYTHIA 8.212 ~\cite{Sjostrand:2014zea}.
The \ttbar process contribution is then assigned a correction based on the top quark \pt, which is known to be improperly modeled~\cite{Khachatryan:2016mnb}.
A correction for the relative fraction of W+light quark/gluon jets and W+charm/bottom jets in W+jets events is derived and then checked in a control region.
More details on the background estimation methods can be found in Section \ref{sec:anstrat}.

All simulated signal and background samples are processed through \PYTHIA for parton fragmentation and hadronization.
The simulation of the CMS detector is performed by \GEANT4~\cite{Agostinelli:2002hh,Allison:2006ve}.
The NNPDF 3.0 parton distribution function (PDF) set is used for sample generation~\cite{nnpdf}.
All simulated samples include additional proton-proton interactions (pileup) and are weighted such that the distribution of the number of interactions in each event agrees with that in the data.
\section{Event selection\label{sec:select}}

All leptons, jets, and \ptvecmiss used in this search are reconstructed using the particle-flow algorithm.
Jets are clustered using the anti-\kt algorithm~\cite{Cacciari:2008gp, Cacciari:2011ma} with a size parameter of 0.4\,(AK4), and dedicated jet energy corrections~\cite{Chatrchyan:2011ds,Khachatryan:2016kdb} are then applied.
Any charged hadrons that are not associated with the leading vertex are removed from the event, using the charged hadron subtraction method~\cite{Krohn:2013lba}.
The leading vertex is defined as the primary vertex with the largest squared sum of the transverse momenta of its associated tracks.
The neutral-hadron contribution to jets from pileup is also subtracted, using the jet area method~\cite{areasubtract}.
Charged hadron subtraction is applied before any jet clustering, while area-based subtractions are applied after clustering but before the final level of jet energy corrections.

Jet momentum is determined as the vectorial sum of all particle momenta in the jet, and is found from simulation to be within 5 to 10\% of the true momentum over the whole \pt spectrum and detector acceptance~\cite{Chatrchyan:2008zzk}.
An offset correction is applied to jet energies to take into account the contribution from pileup.
Jet energy corrections are derived from simulation, and are confirmed with in situ measurements of the energy balance in dijet, multijet, photon+jet, and leptonically decaying Z+jets events.
Additional selection criteria are applied to each event to remove spurious jet-like features originating from isolated noise patterns in certain HCAL regions.

The combined secondary vertex version 2 algorithm~\cite{btagging,CMS:2016kkf} is used to identify jets that have originated from a b quark.
The algorithm combines secondary vertex and track based lifetime information to discriminate b jets from light quark and gluon jets.
The operating point used has a b jet identification (b tagging) efficiency of 80\% and a light-flavor jet misidentification (mistag) probability of 10\%.
Our signal selection requires at least one of the two leading \pt jets to be b-tagged.
This requirement is critical in reducing the contributions from some SM background processes like W+jets.
Scale factors to account for observed differences between data and simulation are applied as a function of \pt.

The event selection, which is optimized separately for the electron and muon channels, results in different requirements for the two channels.
Most notably, the multijet background, through misidentification of showers, is significantly larger in the electron channel than in the muon channel.
For electron events we therefore require higher $\abs{\ptvecmiss}$ and correspondingly lower leading jet \pt than for muon events, in order to keep acceptance high for signal events.

Events are required to have at least two jets with $\pt>30$\GeV and $\abs{\eta}<2.4$, and the leading \pt jet must have $\pt>350\,(450)$\GeV in the electron (muon) channel.

One lepton in each event is required to have fired a single-lepton trigger that has no isolation requirement, be within the detector
acceptance ($\abs{\eta} <2.5$ for electrons, excluding the barrel endcap transition region, $1.444<\abs{\eta}<1.566$, and $\abs{\eta} < 2.4$ for muons) and be associated with a reconstructed primary vertex.
For heavy \PWpr resonance masses, the top quark from the \PWpr decay is highly boosted, causing the b-jet and lepton to be close to each other. For this reason, leptons are not required to be isolated.
Electrons and muons are required to have $\pt > 180$\GeV and to fulfill several identification criteria. Electron candidates are selected using a boosted decision tree based on the shower shape information,
the quality of the track, the match between the track and electromagnetic cluster, the fraction of total cluster energy in the hadronic calorimeter, the amount of activity in the surrounding regions of the
tracker and calorimeters, and the probability of the electron originating from a converted photon. The track associated with a muon candidate is required to have hits in the pixel and muon detectors, a
good-quality fit, and be consistent with originating from the primary vertex.
To reduce the multijet background, the candidate lepton is required to satisfy either $\Delta R$(lepton, nearest jet) $> 0.4$ or $\pt^{\text{rel}}$(lepton, nearest jet) ${>}60\,(50)$\GeV for electrons (muons),
 where $\Delta R=\sqrt{\smash[b]{(\Delta\eta)^2+(\Delta\phi)^2}}$ and $\pt^{\text{rel}}$ is defined as the magnitude of the lepton momentum orthogonal to the jet axis.
Events with additional charged leptons with $\pt > 35$\GeV and $\abs{\eta} < 2.5$ for electrons and $\abs{\eta} < 2.4$ for muons are vetoed.

The four-vectors of identified lepton candidate particles are subtracted from those of jets containing them.
This procedure helps to ensure the reconstructed jets are not contaminated by nearby high-energy leptons as is common in the characteristic boosted signal topology.
Scale factors resulting from small differences between lepton identification and trigger efficiencies in data and simulation are derived in a $\Z\to\ell\ell$ sample as a function of $\abs{\eta}$ and $\pt$ and applied as a correction to simulated events.

Events are required to have at least $\abs{\ptvecmiss}>120\,(50)$\GeV in the electron (muon) channel.
Additionally, events in the electron channel must have $\abs{\Delta\phi(\Pe,\ptvecmiss)}<2$ radians.
These requirements are responsible for differences between the two channels in yields from some background processes.
This selection, along with the other requirements, also helps reject nearly all multijet background events.

\subsection{Mass reconstruction}

The tb invariant mass is reconstructed from the momenta of the charged
lepton and two jets in the event, together with the \ptvecmiss.
The transverse components of the neutrino momentum are set to the \ptvecmiss and the longitudinal component $p^\nu_z$ is calculated by constraining the invariant mass of the
lepton and neutrino to the W boson mass. This method leads to a quadratic equation in $p^\nu_z$.
In the case that the two solutions are real numbers, both solutions are used to reconstruct W boson
candidates. If both solutions contain imaginary parts, then $p^\nu_z$ is set to the real part of the
solutions, and then recompute $\pt^\nu$, which yields another quadratic ambiguity. In this case, we
use only the solution with the mass closest to 80.4\GeV.
Once all the components of the neutrino momentum have been assigned, the viable
solutions for the neutrino are combined with the charged lepton to define W boson candidate(s).
The top quark candidate is then reconstructed by combining the four-momenta of each W boson candidate with each jet with $\pt>25$\GeV and $\abs{\eta}<2.4$.
The jet that yields a top quark mass closest to the nominal top quark mass is used to reconstruct the top quark
candidate. In the case of two W candidates, only the candidate that yields the best top quark mass is used.
Finally, the top quark candidate is combined with the highest \pt jet remaining in the event, yielding the reconstructed $\PWpr$ candidate.
The mass of the $\PWpr$ candidate is referred to as \mtb.

Additional requirements that improve the rejection of background events are placed on the combinations of objects involved in the mass reconstruction.
The top quark candidate is required to have $\pt^{\PQt}>250$\GeV and $100<m_{\PQt}<250$\GeV, and $\pt^\mathrm{j_1+j_2}>350$\GeV, where $\pt^\mathrm{j_1+j_2}$ is the \pt of the four-vector sum of the two leading \pt jets.

Two event categories based on $\pt^{{\PQt}}$ and $\pt^\mathrm{j_1+j_2}$ are used when setting cross section limits.
All events satisfying the above criteria are classified as Type A except for those with $\pt^{{\PQt}}>650$\GeV and $\pt^\mathrm{j_1+j_2}>700$\GeV, which are labeled Type B events.
This categorization improves the sensitivity to high signal masses without sacrificing the performance for lower masses.

Finally, events are also separated into two categories based on whether both (2 b tags) or only one (1 b tag) of the two leading \pt jets is b-tagged.

Event yields in all these categories after the event selection are shown in Table~\ref{table:lep_yields}.

\begin{table*}
\centering
\topcaption{Observed and expected event yields from all the background processes and $\PWpr_\mathrm{R}$ bosons with three different masses. HF and LF indicate heavy flavor and light flavor events, respectively. Yields are separated into eight event categories by the lepton type ($\Pe$ or $\mu$), number of b tags (1 or 2), and $\pt^{\PQt}$ and $\pt^\mathrm{j_1+j_2}$ (Type A or B). The uncertainty in the total expected background includes both the systematic and statistical sources.
\label{table:lep_yields}}
\cmsTable{
\begin{tabular}{lcc{c}@{\hspace*{5pt}}cc{c}@{\hspace*{5pt}}cc{c}@{\hspace*{5pt}}cc}
 & \mc{5}{c}{Electron channel} && \mc{5}{c}{Muon channel} \\ \cline{2-6} \cline{8-12}
 & \mc{2}{c}{Type A} && \mc{2}{c}{Type B} && \mc{2}{c}{Type A} && \mc{2}{c}{Type B}\\ \cline{2-3} \cline{5-6} \cline{8-9} \cline{11-12}
Process & \mc{1}{c}{1 b tag} & \mc{1}{c}{2 b tags} && \mc{1}{c}{1 b tag} & \mc{1}{c}{2 b tags} && \mc{1}{c}{1 b tag} & \mc{1}{c}{2 b tags} && \mc{1}{c}{1 b tag} & \mc{1}{c}{2 b tags}\\[2ex]
{Background} & \\
\ttbar & \mc{1}{c}{760} & \mc{1}{c}{249} && \mc{1}{c}{69} & \mc{1}{c}{22} && \mc{1}{c}{731} & \mc{1}{c}{263} && \mc{1}{c}{75} & \mc{1}{c}{30}\\
tqb & \mc{1}{c}{14} & \mc{1}{c}{6} && \mc{1}{c}{1} & \mc{1}{c}{0} && \mc{1}{c}{14} & \mc{1}{c}{6} && \mc{1}{c}{1} & \mc{1}{c}{0}\\
tW & \mc{1}{c}{117} & \mc{1}{c}{50} && \mc{1}{c}{15} & \mc{1}{c}{5} && \mc{1}{c}{116} & \mc{1}{c}{44} && \mc{1}{c}{22} & \mc{1}{c}{5}\\
tb & \mc{1}{c}{2} & \mc{1}{c}{2} && \mc{1}{c}{0} & \mc{1}{c}{0} && \mc{1}{c}{3} & \mc{1}{c}{1} && \mc{1}{c}{0} & \mc{1}{c}{0}\\
$\PW(\to\ell\nu)$+jets (LF) & \mc{1}{c}{189} & \mc{1}{c}{17} && \mc{1}{c}{16} & \mc{1}{c}{2} && \mc{1}{c}{177} & \mc{1}{c}{16} && \mc{1}{c}{15} & \mc{1}{c}{1}\\
$\PW(\to\ell\nu)$+jets (HF) & \mc{1}{c}{581} & \mc{1}{c}{98} && \mc{1}{c}{52} & \mc{1}{c}{7} && \mc{1}{c}{631} & \mc{1}{c}{107} && \mc{1}{c}{51} & \mc{1}{c}{8}\\
$\Z(\to\ell\ell)$+jets & \mc{1}{c}{19} & \mc{1}{c}{11} && \mc{1}{c}{0} & \mc{1}{c}{0} && \mc{1}{c}{64} & \mc{1}{c}{1} && \mc{1}{c}{20} & \mc{1}{c}{0}\\
VV & \mc{1}{c}{35} & \mc{1}{c}{9} && \mc{1}{c}{2} & \mc{1}{c}{0} && \mc{1}{c}{33} & \mc{1}{c}{1} && \mc{1}{c}{5} & \mc{1}{c}{4}\\[2ex]
{Total background} & \mc{1}{c}{1717$\pm$62} & \mc{1}{c}{442$\pm$34} && \mc{1}{c}{155$\pm$23} & \mc{1}{c}{36$\pm$7} && \mc{1}{c}{1769$\pm$70} & \mc{1}{c}{439$\pm$30} && \mc{1}{c}{189$\pm$22} & \mc{1}{c}{48$\pm$9}\\[2ex]
{Data} & \mc{1}{c}{1750} & \mc{1}{c}{437} && \mc{1}{c}{133} & \mc{1}{c}{40} && \mc{1}{c}{1754} & \mc{1}{c}{482} && \mc{1}{c}{164} & \mc{1}{c}{44}\\[2ex]
{Signal} & \\
$M_{\PWpr_\mathrm{R}}=2000$\GeV & \mc{1}{c}{53} & \mc{1}{c}{43} && \mc{1}{c}{41} & \mc{1}{c}{25} && \mc{1}{c}{79} & \mc{1}{c}{75} && \mc{1}{c}{57} & \mc{1}{c}{35}\\
$M_{\PWpr_\mathrm{R}}=2600$\GeV & \mc{1}{c}{8} & \mc{1}{c}{6} && \mc{1}{c}{16} & \mc{1}{c}{10} && \mc{1}{c}{14} & \mc{1}{c}{12} && \mc{1}{c}{24} & \mc{1}{c}{15}\\
$M_{\PWpr_\mathrm{R}}=3200$\GeV & \mc{1}{c}{2} & \mc{1}{c}{1} && \mc{1}{c}{4} & \mc{1}{c}{3} && \mc{1}{c}{3} & \mc{1}{c}{2} && \mc{1}{c}{8} & \mc{1}{c}{5}\\
\end{tabular}
}
\end{table*}
\section{Backgrounds\label{sec:anstrat}}

\subsection{The W+jets background}

For the W+jets background, the relative fractions of the heavy and light flavor components in simulation are known to differ from those in data~\cite{Chatrchyan:2011yy}.
The validity of the modeling of the flavor content is tested and two scale factors are derived for W+jets heavy and light flavor events using two samples that differ from the signal selection only in b tagging.
The \textit{pre tag} sample does not have any b tagging requirements, while the events in the \textit{0 tag} sample must not have any b-tagged jets.
In these two regions the relative fractions of the W+jets heavy and light flavor events are distinctly different.
The yields from data and simulation in these two regions are used to solve a system of equations for the relative fractions of W+jets heavy and light flavor components, while requiring that the overall W+jets yield remains unchanged.
Uncertainties are determined from repeating the calculation after varying the b tagging efficiencies and mistag rates within their uncertainties.
The scale factors are found to be $2.10\pm^{0.21}_{0.18}$ and $0.49\pm^{0.08}_{0.10}$ for W+jets heavy and light flavor events, respectively.
The corresponding scale factor is then applied to all simulated W+jets events.

\subsection{The top quark pair production background}

For the \ttbar background, we verify normalization as well as the modeling of the top quark \pt. This check is performed in two
signal-depleted \ttbar-enriched regions: one that requires $450<\mtb<750$\GeV and at least two b tags, and another that removes the second-lepton veto and instead requires an additional electron or muon with a \pt of at least 35\GeV.
These comparisons motivate a reweighting of the \ttbar background using a correction factor obtained from measurements of the differential top quark \pt distribution.
This correction factor is applied to the \ttbar simulation, as a function of the generator-level top quark \pt.
The \ttbar simulation without the correction factor applied is used as an estimate of the systematic uncertainty in the reweighting procedure.
\section{Systematic uncertainties\label{sec:syst}}

The systematic uncertainties in this analysis can be grouped into two categories: uncertainties in the overall normalization and in the
shape of the \mtb distribution.

The normalization uncertainties include the uncertainty in the integrated luminosity (2.5\%)~\cite{CMS-PAS-LUM-17-001}, the \ttbar and W+jets cross sections (8 and 10\%, respectively), the lepton identification (2\%), and the trigger efficiencies (2\%).

The uncertainty due to variations in the renormalization and factorization scales ($\mu_\mathrm{R}$ and $\mu_\mathrm{F}$, respectively) is evaluated at the matrix element level using event weights from varying the scales by 0.5 and 2 while restricting to $0.5\le\mu_\mathrm{R}/\mu_\mathrm{F}\le2$~\cite{Cacciari:2003fi,Catani:2003zt}.

Uncertainties resulting from $\pm1$ standard deviation (s.d.) variations in the b tagging efficiency and mistagging rate scale factors, jet energy scale, and jet energy resolution are also included.

A correction is applied to all simulated samples to better match the distribution of pileup interactions observed in data.
This procedure uses a total inelastic cross section of 69.2\unit{mb}, and an uncertainty is calculated by varying the cross section by $\pm$5\%~\cite{Aaboud:2016mmw}.

To estimate the uncertainty arising from the choice of PDF, we evaluate the root-mean-square of the distribution of 100 NNPDF 3.0 replicas as the $\pm$1 s.d. uncertainties according to the guidelines in Ref.~\cite{pdf4lhc}.
When considering signal samples only the shape component of the uncertainty due to PDFs is included.

The uncertainty in the W+jets heavy and light flavor scale factors is included as a variation in the W+jets background.
The \ttbar background with an uncorrected top quark \pt spectrum is included as a one-sided $+1$ s.d. variation.

All uncertainties are listed in Table~\ref{table:systematics}.
The uncertainties with the largest effect on the overall background normalization are those associated with the top quark \pt reweighting, $\mu_\mathrm{R}$ and $\mu_\mathrm{F}$ scales, and PDFs, which have effects of approximately 15, 15, and 6\%, respectively.

\begin{table}
\centering
\topcaption{List of systematic uncertainties taken into account in the analysis. For sources that affect the shape of the \mtb distribution the given rate uncertainty is approximate. The pileup, top quark \pt reweighting, and W+jets heavy/light flavor systematic uncertainties are described in more detail in the text. A check mark in the ``Signal" column indicates that the uncertainty is also applied to the signal samples. For the PDF uncertainty, only its shape component is included for signal samples.}
\label{table:systematics}
\begin{tabular}{lcc}
Source & Rate uncertainty & Signal \\\hline
\multicolumn{1}{l}{{Normalization}} & &\\
Integrated luminosity & 2.5\% & \checkmark \\
$\ttbar$ cross section & 8\% & --- \\
W+jets cross section & 10\% & --- \\
Trigger eff. ($\Pe/\mu$) & 2\%/2\% & \checkmark \\
Lepton id. eff. ($\Pe/\mu$) & 2\%/2\% & \checkmark \\[2ex]
\multicolumn{1}{l}{{Shape and normalization}} & & \\
Jet energy scale & 3\% & \checkmark \\
Jet energy resolution & 1\% & \checkmark \\
b/c tagging & 2\% & \checkmark \\
Light quark mistagging & 2\% & \checkmark \\
Pileup & 1\% & \checkmark \\
PDF & 6\% & \checkmark \\
Top quark \pt reweighting & 15\% & --- \\
W+jets heavy/light flavor & 1\% & --- \\
$\mu_\mathrm{R}$ and $\mu_\mathrm{F}$ scales & 15\% & --- \\
\end{tabular}
\end{table}
\section{Results\label{sec:results}}

\begin{figure*}
  \centering
    \includegraphics[width=0.48\textwidth]{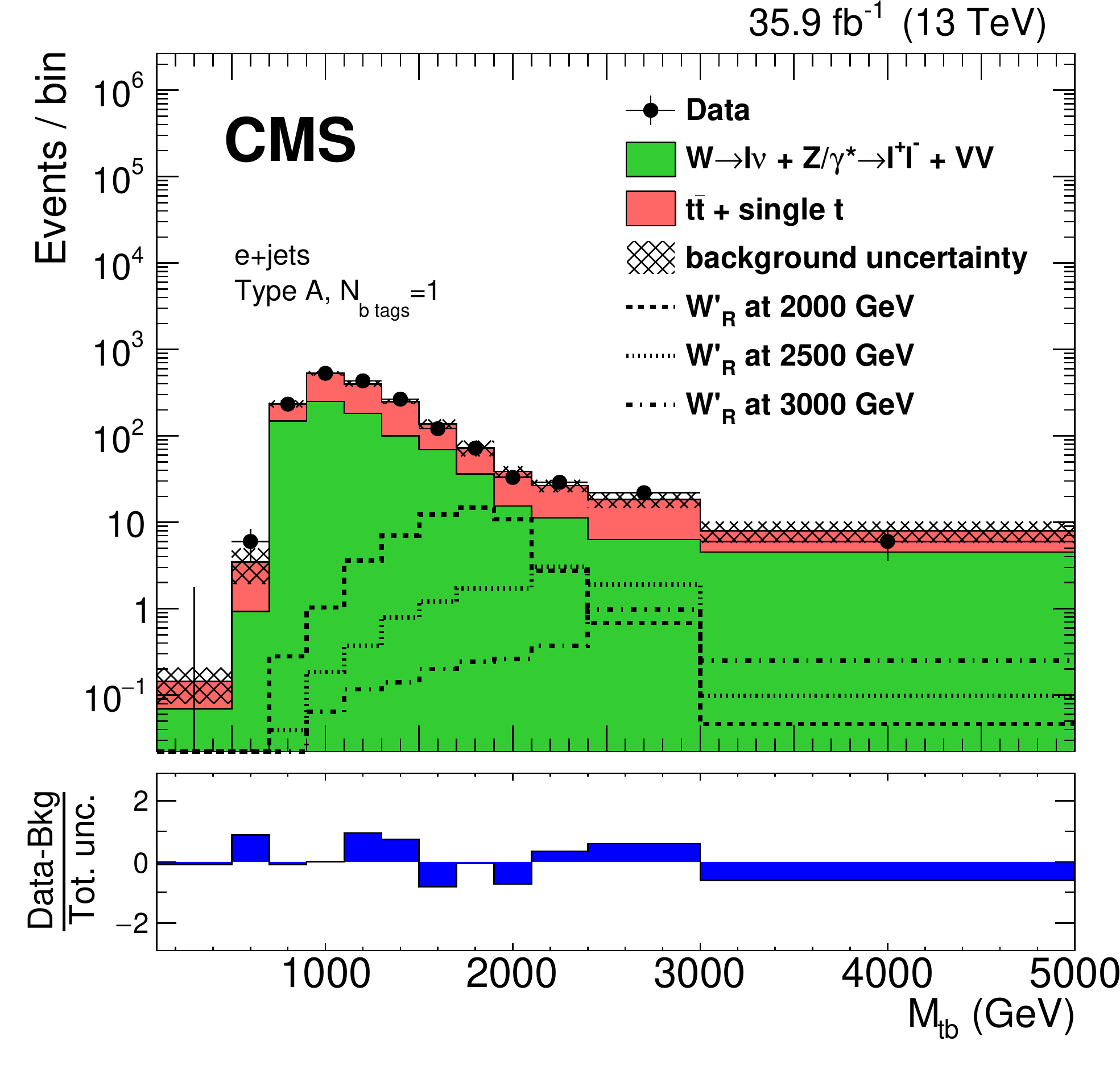}
    \includegraphics[width=0.48\textwidth]{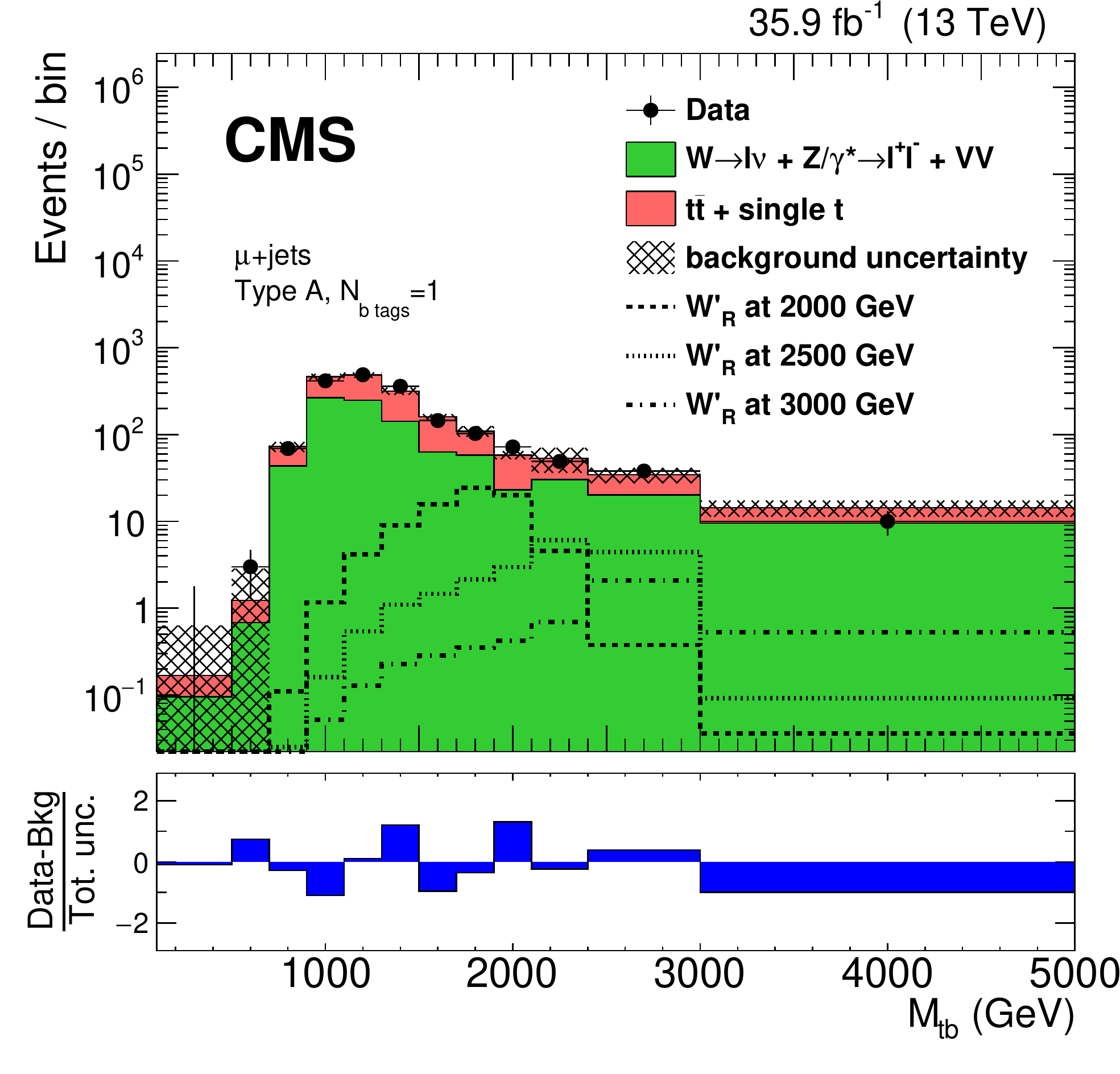}
    \includegraphics[width=0.48\textwidth]{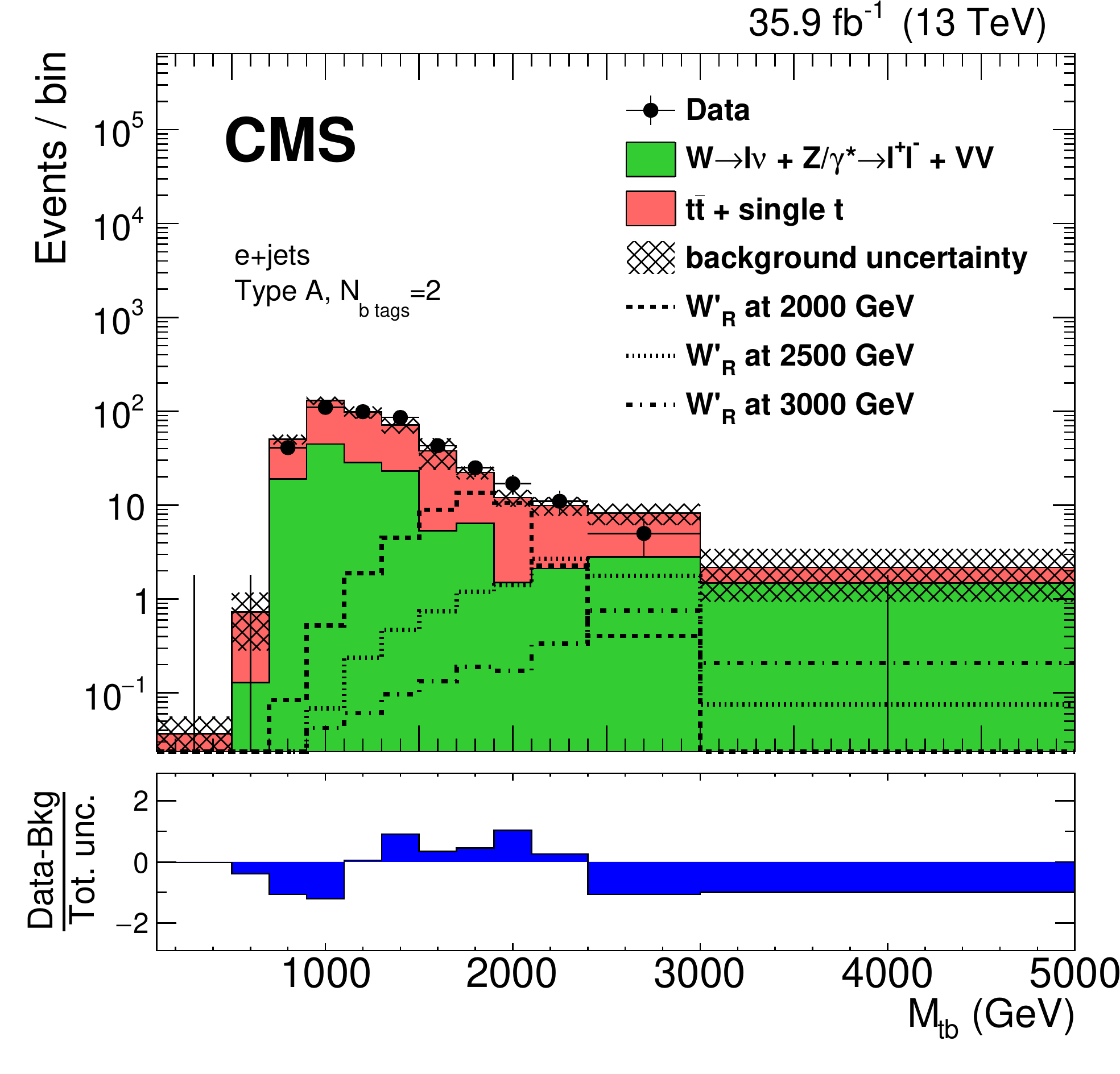}
    \includegraphics[width=0.48\textwidth]{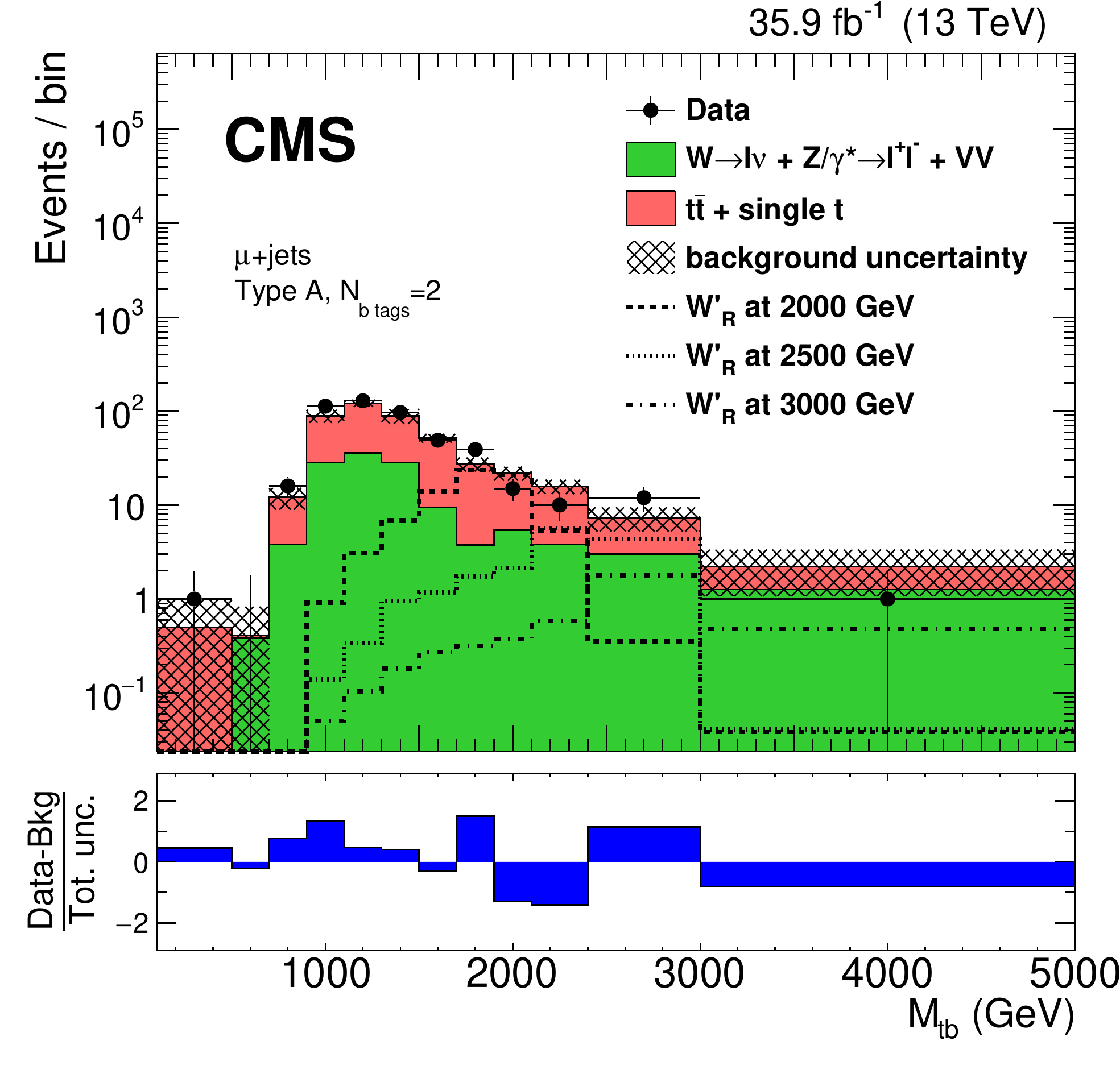}
    \caption{The reconstructed \mtb distributions in the 1 b tag (upper) and 2 b tags (lower) categories, for the electron (left) and muon (right) channels, for Type A events.
Distributions for $\PWpr_\mathrm{R}$ bosons with masses of 2, 2.5, and 3\TeV are shown.
The distribution is shown after the application of all selections.
The background uncertainty includes both statistical and systematic components, while ``Tot. unc." in the lower panels corresponds to the combined uncertainty of the background prediction and data.
}
 \label{fig:lep_mtb_A}
\end{figure*}

\begin{figure*}
  \centering
    \includegraphics[width=0.48\textwidth]{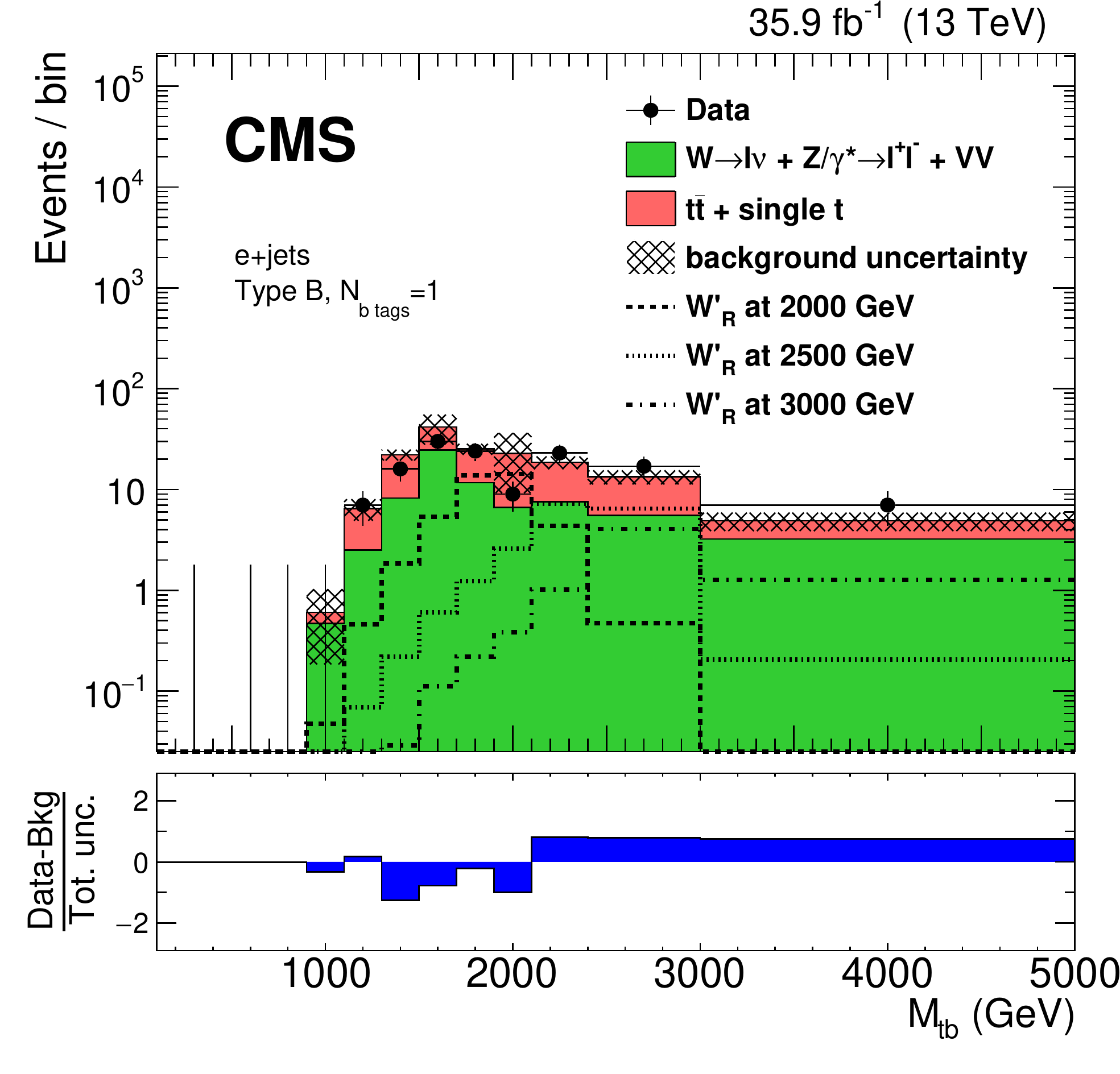}
    \includegraphics[width=0.48\textwidth]{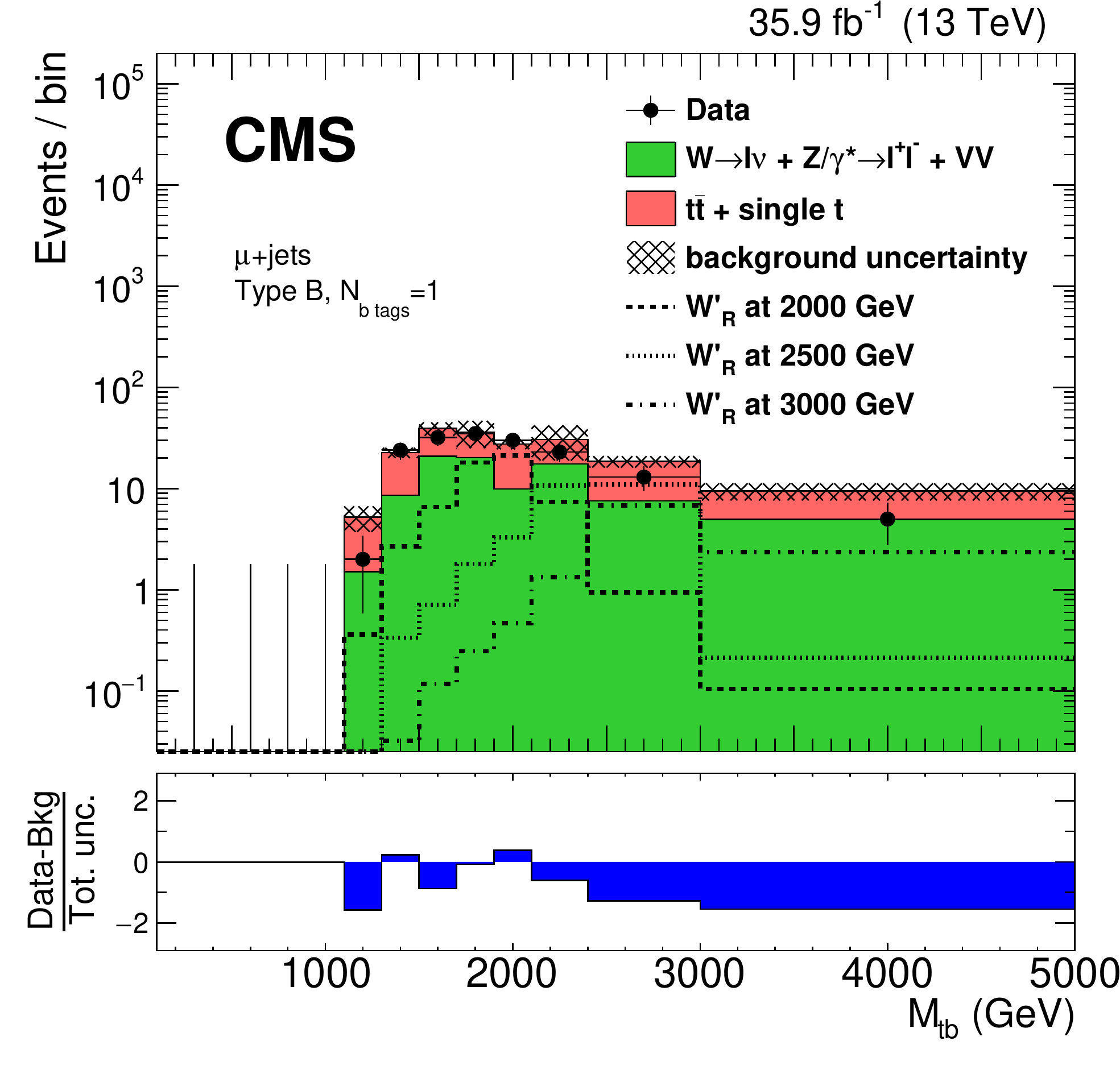}
    \includegraphics[width=0.48\textwidth]{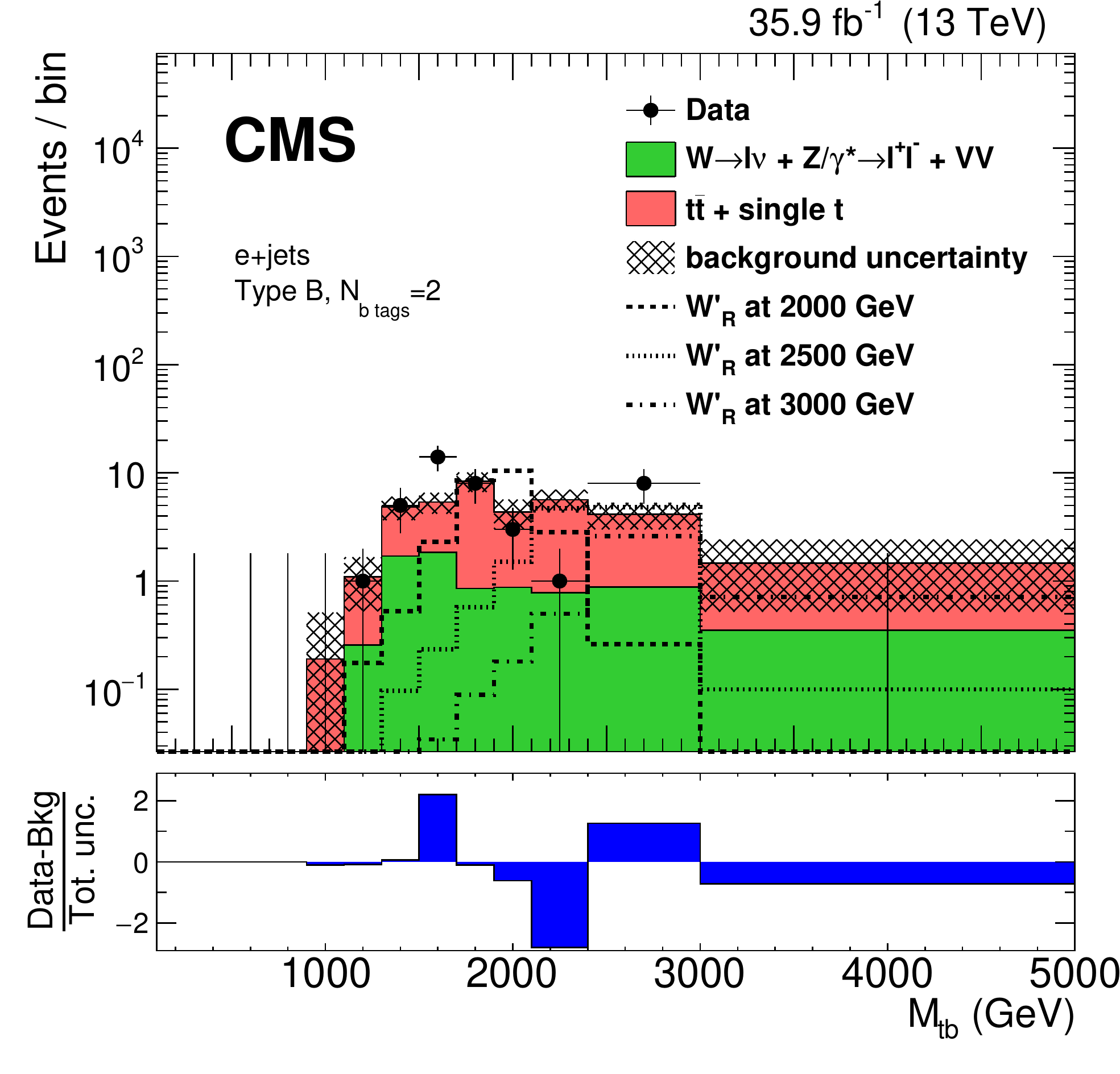}
    \includegraphics[width=0.48\textwidth]{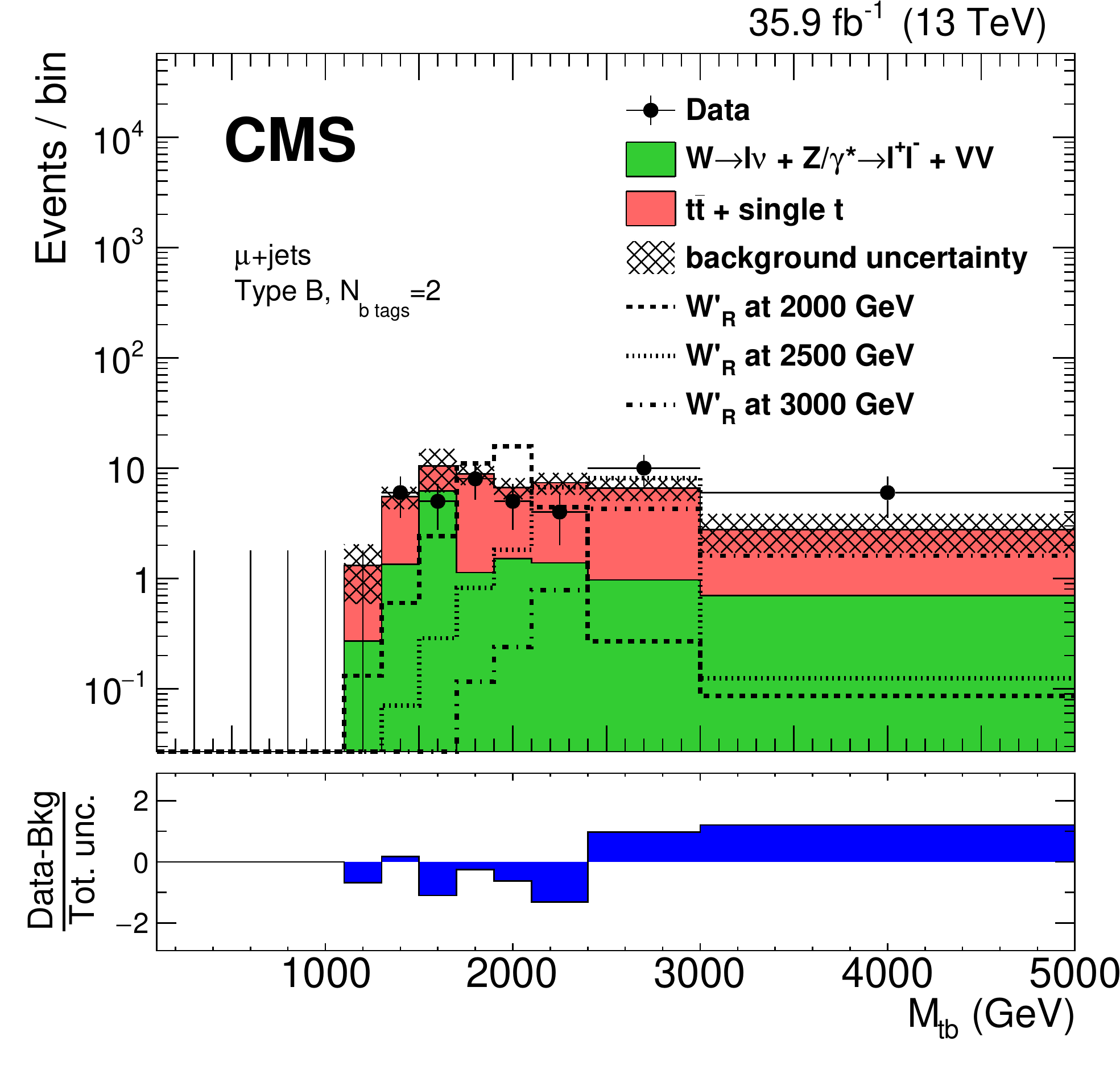}
    \caption{The reconstructed \mtb distributions in the 1 b tag (upper) and 2 b tags (lower) categories, for the electron (left) and muon (right) channels, for Type B events.
Distributions for $\PWpr_\mathrm{R}$ bosons with masses of 2, 2.5, and 3\TeV are shown.
The distribution is shown after the application of all selections.
The background uncertainty includes both statistical and systematic components, while ``Tot. unc." in the lower panels corresponds to the combined uncertainty of the background prediction and data.
}
\label{fig:lep_mtb_B}
\end{figure*}

Distributions of \mtb are shown in Figs.~\ref{fig:lep_mtb_A} and~\ref{fig:lep_mtb_B}.
The binning is chosen to reduce uncertainties due to the size of the simulated event samples and is one bin from 0 to 500\GeV, eight bins of 200\GeV width from 500 to 2100\GeV, one bin from 2100 to 2400\GeV, one bin from 2400 to 3000\GeV, and one bin above 3000\GeV.
Having observed that data agree with the predicted SM background processes, we set 95\% confidence level (CL) upper limits on the $\PWpr$ boson production cross section for masses between 1 and 4\TeV.

The analysis separates events into eight independent categories in order to improve the signal sensitivity.
Categories are created according to lepton type (electron or muon), the number of b-tagged jets among the first two leading \pt jets (1 or 2), and $\pt^{\PQt}$ and $\pt^\mathrm{j_1+j_2}$ (Type A or B).
Categorization according to the number of b tags allows the analysis to maintain acceptance for signal events where one of the jets is not correctly b tagged, and categorization according to the $\pt^{\PQt}$ and $\pt^\mathrm{j_1+j_2}$ allows the analysis to perform well over a large range of possible signal masses.

Limits on the cross section of $\PWpr$ bosons are calculated using a Bayesian method with a prior uniform in the signal cross section, as implemented with the \textsc{theta} package~\cite{theta-stat}.
The Bayesian approach uses a binned likelihood in order to calculate the 95\% CL upper limits on the product of the signal production and the branching fraction $\sigma(\Pp\Pp\to{\PWpr})\,\mathcal{B}(\PWpr\to\PQt\PQb)$.
Statistical uncertainties related to the background prediction are treated using the ``Barlow--Beeston lite" method~\cite{barlowbeeston}.
All uncertainties given in Section~\ref{sec:syst} are included as nuisance parameters.
Uncertainties in the shape of the \mtb distribution are treated using template interpolation and all rate uncertainties are included with log-normal priors.

\begin{figure}[htb]
 \centering
  \includegraphics[width=0.48\textwidth]{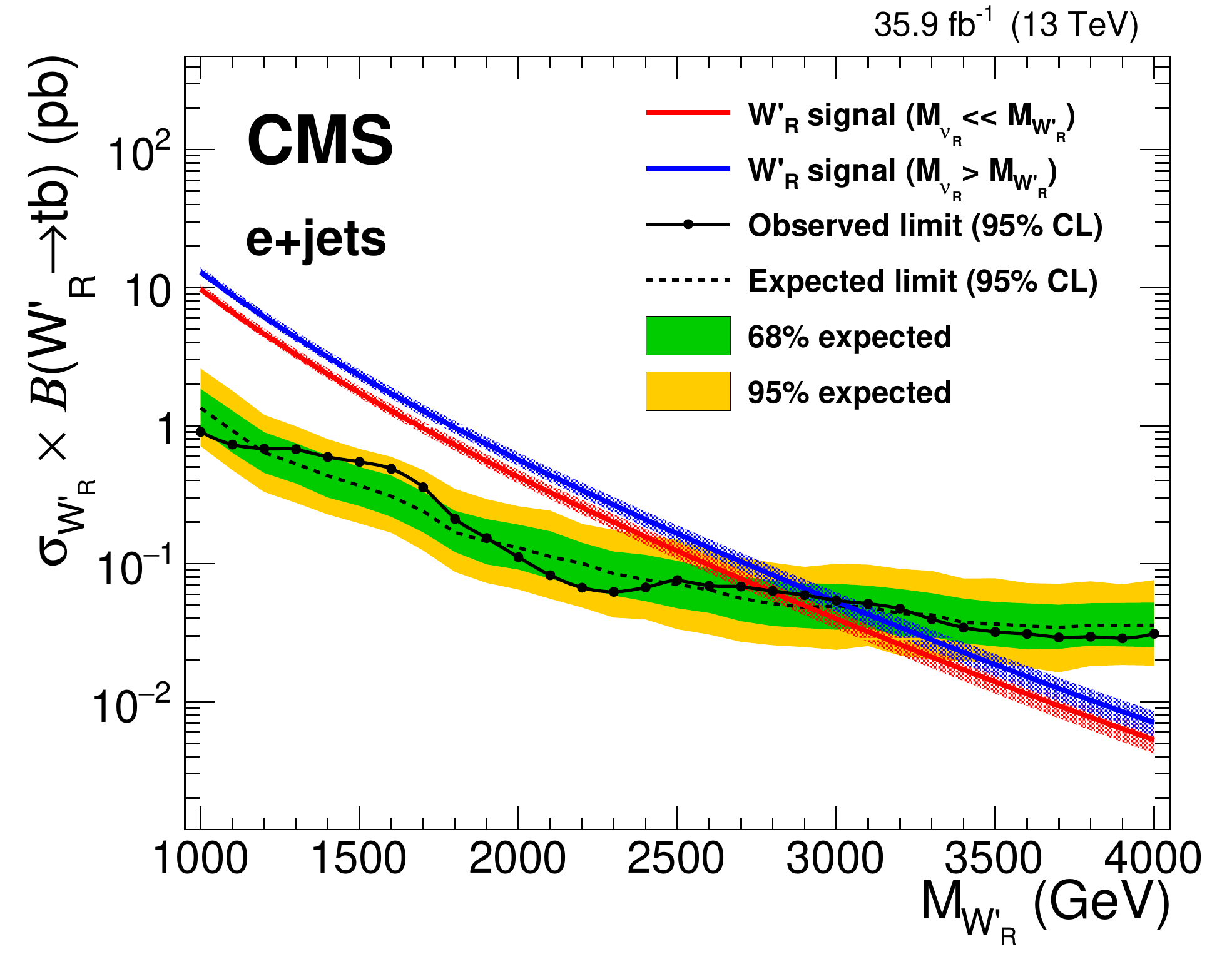}
  \includegraphics[width=0.48\textwidth]{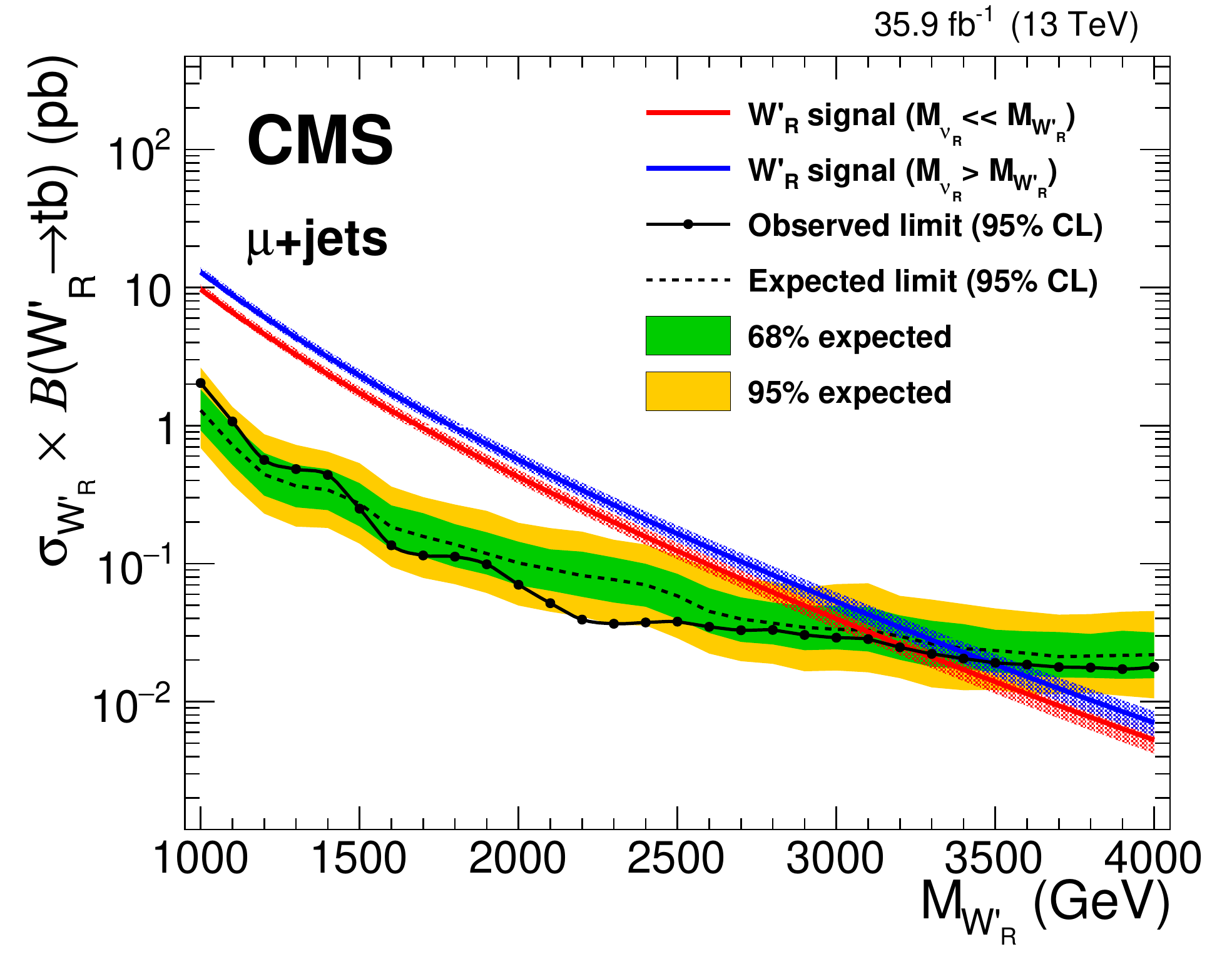}
 \caption{Upper limit at 95\% CL on the $\PWpr_\mathrm{R}$ boson production cross section separately in the electron (\cmsLeft) and muon (\cmsRight) channels. Signal masses for which the theoretical cross section (in red and blue for $M_{\nu_\mathrm{R}}\ll M_{\PWpr_\mathrm{R}}$ and $M_{\nu_\mathrm{R}}>M_{\PWpr_\mathrm{R}}$, respectively) exceeds the observed upper limit (in solid black) are excluded at 95\% CL.
 The green and yellow bands represent the $\pm$1 and 2 s.d. uncertainties in the expected limit, respectively.\label{fig:sepxseclim_sep}}
\end{figure}

\begin{figure}
 \centering
  \includegraphics[width=0.48\textwidth]{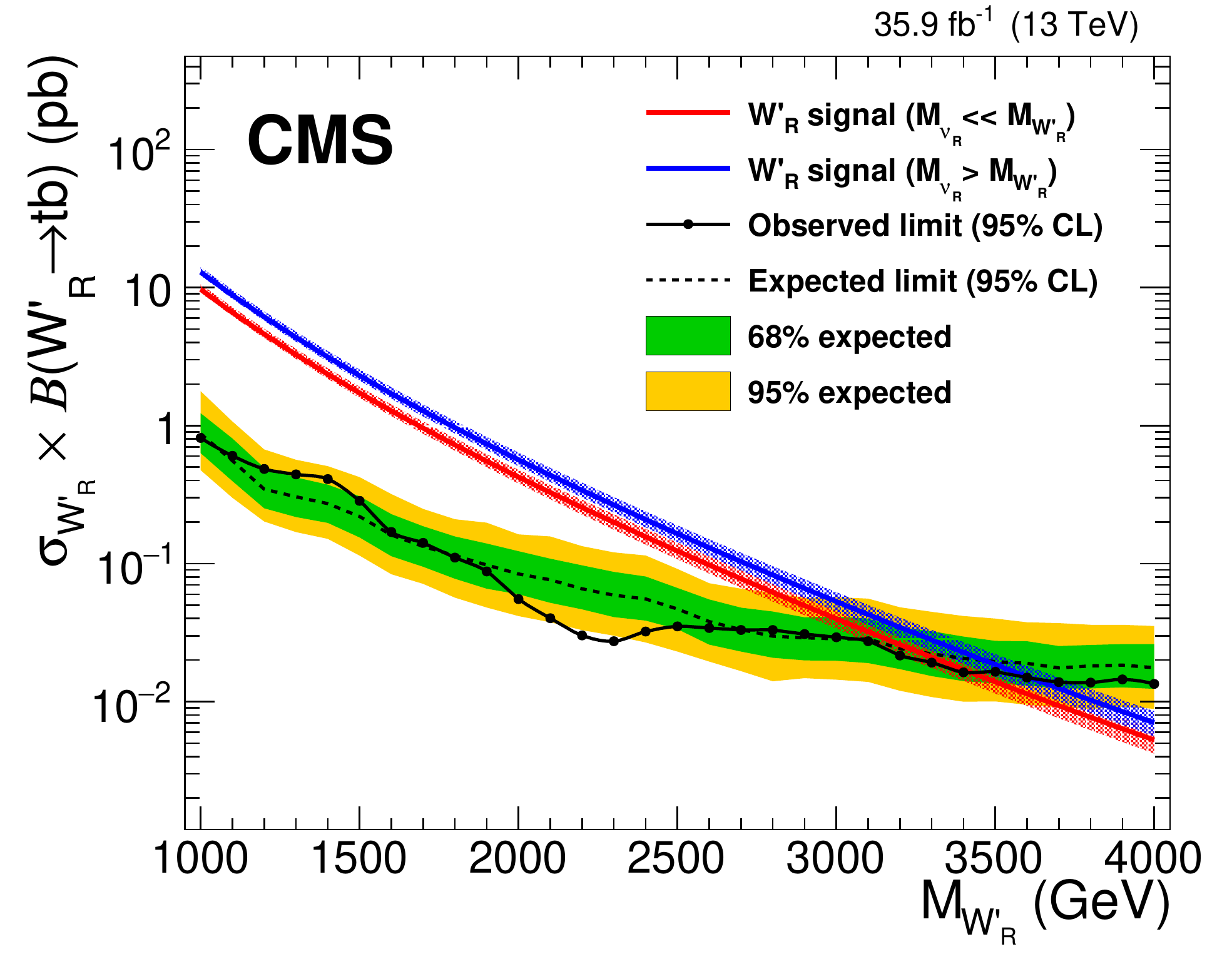}
 \caption{Upper limit at 95\% CL on the $\PWpr_\mathrm{R}$ boson production cross section for the combined electron and muon channels. Signal masses for which the theoretical cross section (in red and blue for $M_{\nu_\mathrm{R}}\ll M_{\PWpr_\mathrm{R}}$ and $M_{\nu_\mathrm{R}}>M_{\PWpr_\mathrm{R}}$, respectively) exceeds the observed upper limit (in solid black) are excluded at 95\% CL.
 The green and yellow bands represent the $\pm$1 and 2 s.d. uncertainties in the expected limit, respectively.\label{fig:sepxseclim}}
\end{figure}

Results for right-handed \PWpr bosons are shown in Figs.~\ref{fig:sepxseclim_sep} and~\ref{fig:sepxseclim}.
$\PWpr_\mathrm{R}$ bosons with masses below 3.4\TeV are excluded at 95\% CL.

Although models with a \PWpr boson that couples exclusively to right-handed fermions are simpler because of the lack of interference, the effective Lagrangian in Eq.~(\ref{eq:lag}) allows us to analyze models with arbitrary combinations of left- and right-handed couplings.
In order to accomplish this the
interference between the SM $s$-channel tb production and the tb production via an intermediate left-handed W' boson must be accounted for since these processes initial and final states are identical.

The cross section for single top quark production given a \PWpr boson can be written for any set of $a_\mathrm{L}$ and $a_\mathrm{R}$ coupling values in terms of the cross sections of four simulated signal samples.
It is assumed that the couplings to fermions are independent of generation, such that each signal can be described by a single value of $a_\mathrm{L}$ and a single value of $a_\mathrm{R}$.
The four simulated signals are then
$\sigma_\mathrm{L}$ for purely left-handed couplings $(a_\mathrm{L},a_\mathrm{R})=(1,0)$,
$\sigma_\mathrm{R}$ for purely right-handed couplings $(a_\mathrm{L},a_\mathrm{R})=(0,1)$,
$\sigma_\mathrm{LR}$ for mixed couplings $(a_\mathrm{L},a_\mathrm{R})=(1/\sqrt{2},1/\sqrt{2})$,
and $\sigma_\mathrm{SM}$ for SM couplings $(a_\mathrm{L},a_\mathrm{R})=(0,0)$,
and the cross section for single top quark production is
\ifthenelse{\boolean{cms@external}}{
\begin{multline}
 \sigma=(1-a_\mathrm{L}^2)\sigma_\mathrm{SM}+\frac{1}{a_\mathrm{L}^2+a_\mathrm{R}^2}\Bigl[a_\mathrm{L}^2(a_\mathrm{L}^2-a_\mathrm{R}^2)\sigma_\mathrm{L}\\
 +a_\mathrm{R}^2(a_\mathrm{R}^2-a_\mathrm{L}^2)\sigma_\mathrm{R}+4a_\mathrm{L}^2a_\mathrm{R}^2\sigma_\mathrm{LR}-2a_\mathrm{L}^2a_\mathrm{R}^2\sigma_\mathrm{SM}\Bigr].
\end{multline}
}{
\begin{equation}
  \sigma=(1-a_\mathrm{L}^2)\sigma_\mathrm{SM}+\frac{1}{a_\mathrm{L}^2+a_\mathrm{R}^2}\Bigl[a_\mathrm{L}^2(a_\mathrm{L}^2-a_\mathrm{R}^2)\sigma_\mathrm{L}+a_\mathrm{R}^2(a_\mathrm{R}^2-a_\mathrm{L}^2)\sigma_\mathrm{R}+4a_\mathrm{L}^2a_\mathrm{R}^2\sigma_\mathrm{LR}-2a_\mathrm{L}^2a_\mathrm{R}^2\sigma_\mathrm{SM}\Bigr].
\end{equation}
}
By combining four signal samples according to this equation we are able to produce invariant mass distributions for a \PWpr boson with arbitrary $a_\mathrm{L}$ and $a_\mathrm{R}$ couplings.
A notable adjustment for this paper with respect to previous CMS publications is in the definition of the mixed coupling sample, which was previously defined as $(a_\mathrm{L},a_\mathrm{R})=(1,1)$.
This change results in slightly different expressions for the total cross section, and is chosen to ensure that the widths of all three simulated signal samples are identical.

It should be noted that in the case that the \PWpr boson couples exclusively to right-handed fermions, this equation reduces to the sum of SM $s$-channel tb production and $\PWpr_\mathrm{R}$ production, as expected.
For pure $\PWpr_\mathrm{L}$ or $\PWpr_\mathrm{LR}$ boson production, the equation reduces to the cross section of the respective sample, which is generated already including SM $s$-channel tb production and interference with $\PWpr$ production.

\begin{figure}[htbp]
 \centering
  \includegraphics[width=0.48\textwidth]{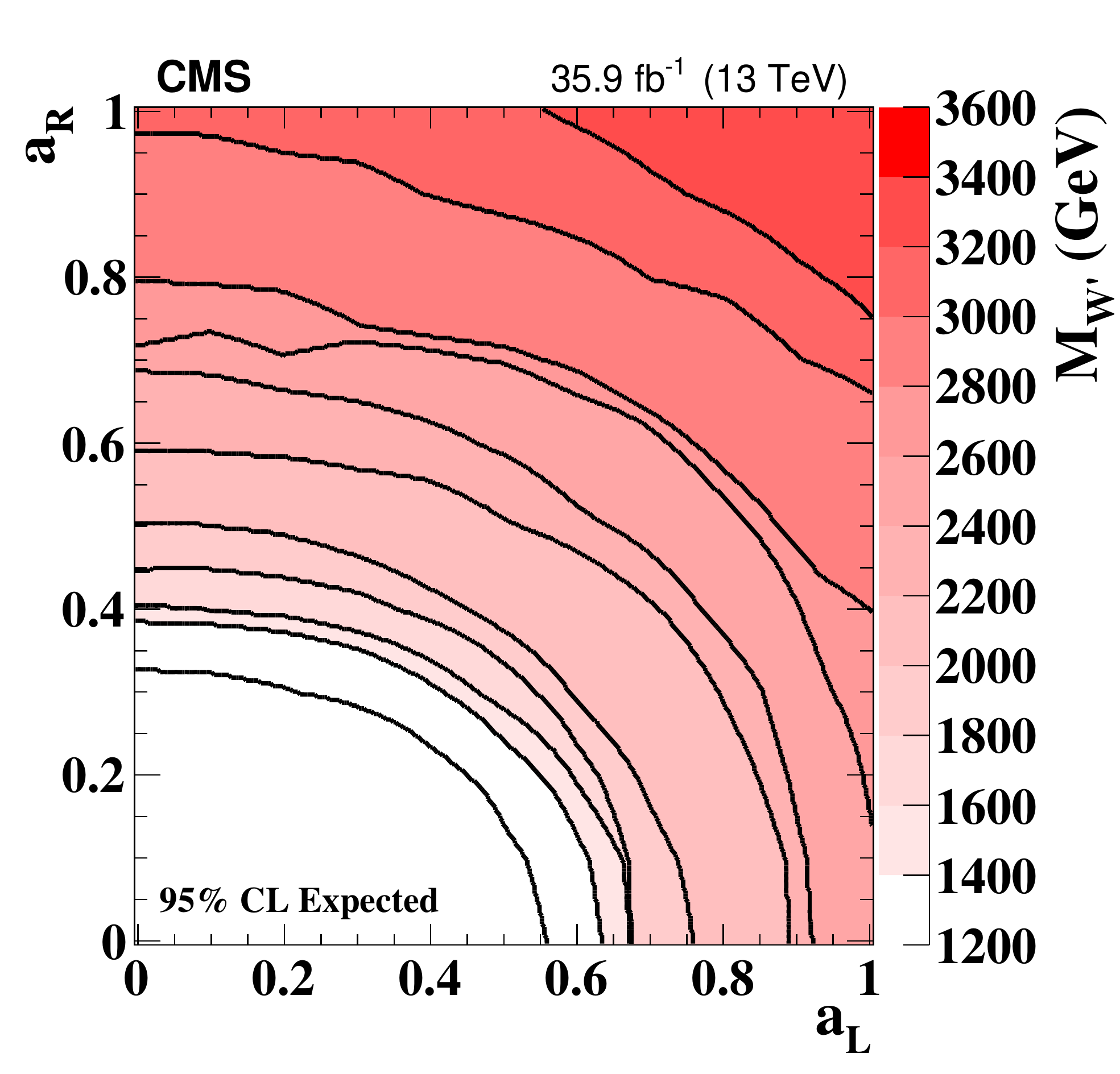}
  \includegraphics[width=0.48\textwidth]{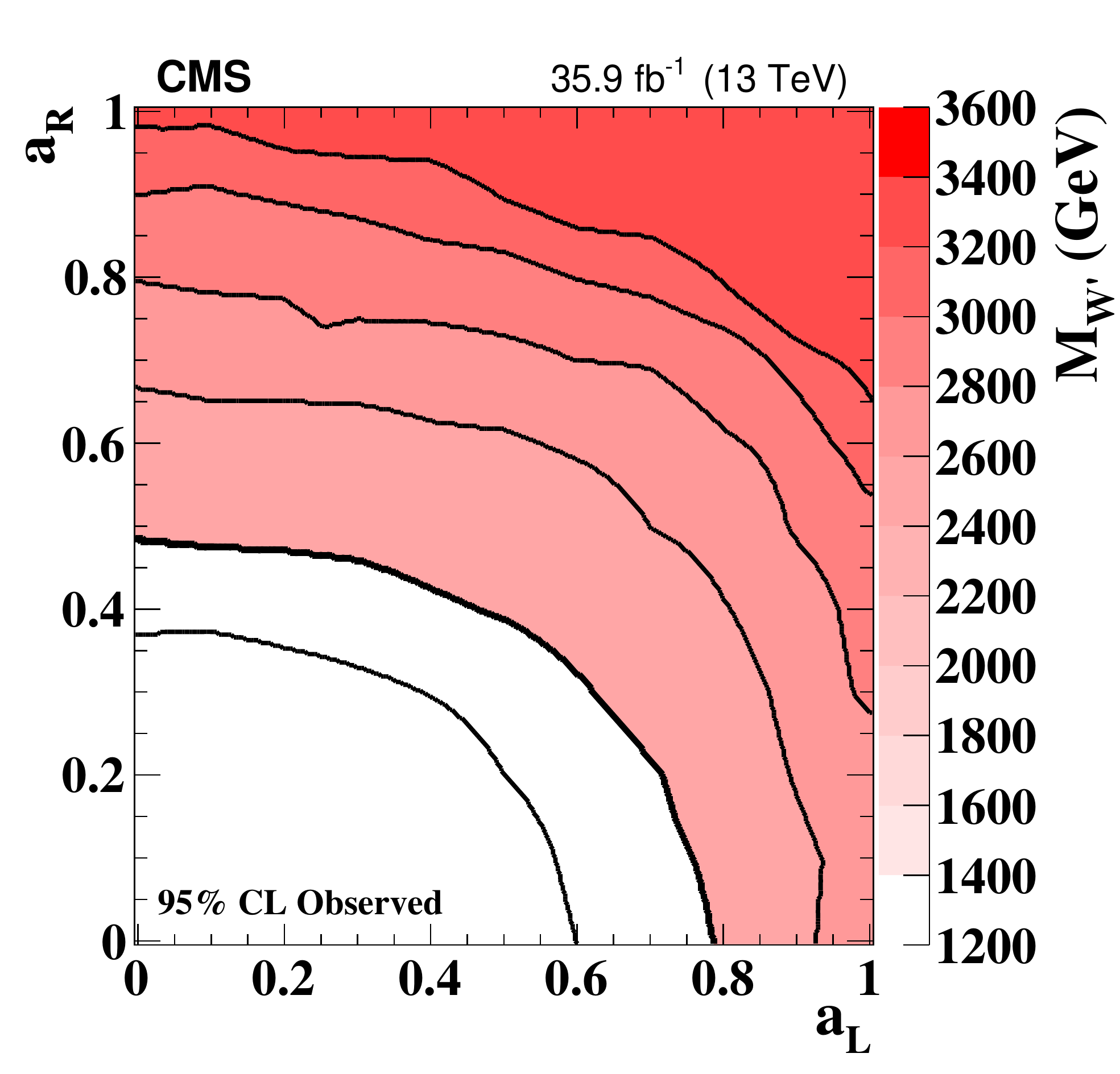}
    \caption{Expected (\cmsLeft) and observed (\cmsRight) limits on the \PWpr boson mass as function of the left-handed ($a_\mathrm{L}$) and right-handed ($a_\mathrm{R}$) couplings.
Black lines represent contours of equal \PWpr boson mass separated by 200\GeV.
\label{fig:2dlimit}}
\end{figure}

A scan is performed over the $a_\mathrm{L}$ and $a_\mathrm{R}$ plane in 0.1 steps from 0 to 1 to produce cross section limits for arbitrary combinations of $a_\mathrm{L}$ and $a_\mathrm{R}$.
For each point in the scan the expected and observed 95\% CL upper limits on the cross section are calculated using the same method described above.
Figure~\ref{fig:2dlimit} shows the excluded \PWpr boson mass for each ($a_\mathrm{L}$, $a_\mathrm{R}$) point, in addition to an interpolation between points to create smooth contours of equivalent signal mass limits.

\section{Summary\label{sec:summary}}

A search for a narrow heavy $\PWpr$ boson resonance decaying to a top quark and a bottom quark has been performed in $\mbox{lepton+jets}$ final states using data collected at $\sqrt{s}=13$\TeV by the CMS detector in 2016, corresponding to an integrated luminosity of 35.9\fbinv.
No evidence is observed for the production of a $\PWpr$ boson, and 95\% CL upper limits on
the product of the right-handed $\PWpr$ ($\PWpr_\mathrm{R}$) boson production cross section and its branching fraction to a top and a bottom quark are calculated as a function of the $\PWpr_\mathrm{R}$ boson mass.
The observed (expected) 95\% CL upper limit is 3.4 (3.3)\TeV if $M_{\PWpr_\mathrm{R}}\gg M_{\nu_\mathrm{R}}$ and 3.6 (3.5)\TeV if $M_{\PWpr_\mathrm{R}}<M_{\nu_\mathrm{R}}$, where $M_{\nu_\mathrm{R}}$ is the mass of the right-handed neutrino.
Exclusion limits are also presented for $\PWpr$ bosons with varied left- and right-handed couplings to fermions, for the first time at $\sqrt{s}=13$\TeV.
These results are the most stringent limits to date on the production of \PWpr bosons that decay to a top and a bottom quark.

\begin{acknowledgments}
We congratulate our colleagues in the CERN accelerator departments for the excellent performance of the LHC and thank the technical and administrative staffs at CERN and at other CMS institutes for their contributions to the success of the CMS effort. In addition, we gratefully acknowledge the computing centres and personnel of the Worldwide LHC Computing Grid for delivering so effectively the computing infrastructure essential to our analyses. Finally, we acknowledge the enduring support for the construction and operation of the LHC and the CMS detector provided by the following funding agencies: BMWFW and FWF (Austria); FNRS and FWO (Belgium); CNPq, CAPES, FAPERJ, and FAPESP (Brazil); MES (Bulgaria); CERN; CAS, MoST, and NSFC (China); COLCIENCIAS (Colombia); MSES and CSF (Croatia); RPF (Cyprus); SENESCYT (Ecuador); MoER, ERC IUT, and ERDF (Estonia); Academy of Finland, MEC, and HIP (Finland); CEA and CNRS/IN2P3 (France); BMBF, DFG, and HGF (Germany); GSRT (Greece); OTKA and NIH (Hungary); DAE and DST (India); IPM (Iran); SFI (Ireland); INFN (Italy); MSIP and NRF (Republic of Korea); LAS (Lithuania); MOE and UM (Malaysia); BUAP, CINVESTAV, CONACYT, LNS, SEP, and UASLP-FAI (Mexico); MBIE (New Zealand); PAEC (Pakistan); MSHE and NSC (Poland); FCT (Portugal); JINR (Dubna); MON, RosAtom, RAS, RFBR and RAEP (Russia); MESTD (Serbia); SEIDI, CPAN, PCTI and FEDER (Spain); Swiss Funding Agencies (Switzerland); MST (Taipei); ThEPCenter, IPST, STAR, and NSTDA (Thailand); TUBITAK and TAEK (Turkey); NASU and SFFR (Ukraine); STFC (United Kingdom); DOE and NSF (USA).

\hyphenation{Rachada-pisek} Individuals have received support from the Marie-Curie programme and the European Research Council and Horizon 2020 Grant, contract No. 675440 (European Union); the Leventis Foundation; the A. P. Sloan Foundation; the Alexander von Humboldt Foundation; the Belgian Federal Science Policy Office; the Fonds pour la Formation \`a la Recherche dans l'Industrie et dans l'Agriculture (FRIA-Belgium); the Agentschap voor Innovatie door Wetenschap en Technologie (IWT-Belgium); the Ministry of Education, Youth and Sports (MEYS) of the Czech Republic; the Council of Science and Industrial Research, India; the HOMING PLUS programme of the Foundation for Polish Science, cofinanced from European Union, Regional Development Fund, the Mobility Plus programme of the Ministry of Science and Higher Education, the National Science Center (Poland), contracts Harmonia 2014/14/M/ST2/00428, Opus 2014/13/B/ST2/02543, 2014/15/B/ST2/03998, and 2015/19/B/ST2/02861, Sonata-bis 2012/07/E/ST2/01406; the National Priorities Research Program by Qatar National Research Fund; the Programa Clar\'in-COFUND del Principado de Asturias; the Thalis and Aristeia programmes cofinanced by EU-ESF and the Greek NSRF; the Rachadapisek Sompot Fund for Postdoctoral Fellowship, Chulalongkorn University and the Chulalongkorn Academic into Its 2nd Century Project Advancement Project (Thailand); and the Welch Foundation, contract C-1845. \end{acknowledgments}
\clearpage
\bibliography{auto_generated}
\cleardoublepage \appendix\section{The CMS Collaboration \label{app:collab}}\begin{sloppypar}\hyphenpenalty=5000\widowpenalty=500\clubpenalty=5000\textbf{Yerevan Physics Institute,  Yerevan,  Armenia}\\*[0pt]
A.M.~Sirunyan, A.~Tumasyan
\vskip\cmsinstskip
\textbf{Institut f\"{u}r Hochenergiephysik,  Wien,  Austria}\\*[0pt]
W.~Adam, F.~Ambrogi, E.~Asilar, T.~Bergauer, J.~Brandstetter, E.~Brondolin, M.~Dragicevic, J.~Er\"{o}, M.~Flechl, M.~Friedl, R.~Fr\"{u}hwirth\cmsAuthorMark{1}, V.M.~Ghete, J.~Grossmann, J.~Hrubec, M.~Jeitler\cmsAuthorMark{1}, A.~K\"{o}nig, N.~Krammer, I.~Kr\"{a}tschmer, D.~Liko, T.~Madlener, I.~Mikulec, E.~Pree, D.~Rabady, N.~Rad, H.~Rohringer, J.~Schieck\cmsAuthorMark{1}, R.~Sch\"{o}fbeck, M.~Spanring, D.~Spitzbart, W.~Waltenberger, J.~Wittmann, C.-E.~Wulz\cmsAuthorMark{1}, M.~Zarucki
% ...
\vskip\cmsinstskip
% ...
1:~~Also at Vienna University of Technology, Vienna, Austria\\
% ...

\end{sloppypar}
\end{document}